\documentclass[12pt]{article}
\usepackage{epsf,epsfig,here,cite}
\newcommand{\GeV}{\rm{GeV}}
\newcommand{\beq}{\begin{equation}}
\newcommand{\eeq}{\end{equation}}
\newcommand{\beqn}{\begin{eqnarray}}  
\newcommand{\eeqn}{\end{eqnarray}}
\newcommand{\bea}{\begin{eqnarray}}
\newcommand{\eea}{\end{eqnarray}}

\newcommand{\rhobar}{\bar {\rho}}
\newcommand{\etabar}{\bar{\eta}}
\newcommand{\epsilonk}{\varepsilon_K}

\newcommand{\snb}{\sin 2\beta}

\newcommand{\fbdsqbd}{f_{B_d} \sqrt{\hat B_{B_d}}}
\newcommand{\fbssqbs}{f_{B_s} \sqrt{\hat B_{B_s}}}
\newcommand{\vcb}{\left | {V_{cb}} \right |}
\newcommand{\vub}{\left | {V_{ub}} \right |}
\def\utfit{{\bf{U}}\kern-.24em{\bf{T}}\kern-.21em{\it{fit}}\@}
\def\utangles{{\bf{U}}\kern-.24em{\bf{T}}\kern-.21em{\it{angles}}\@}
\def\utlattice{{\bf{U}}\kern-.24em{\bf{T}}\kern-.21em{\it{lattice}}\@}

\begin{document}
\pagestyle{empty}
\pagenumbering{arabic}
%\vskip  1.5 cm
\begin{center}
  \begin{LARGE}
    \textbf{The Unitarity Triangle Fit in the \\ Standard Model and  Hadronic Parameters \\ from
      Lattice QCD: \\ 
      A Reappraisal after the Measurements \\
      of $\Delta m_s$ and $BR(B\to \tau \nu_\tau)$} \\
  \end{LARGE}  
\end{center}

\vspace*{-0.2cm}
\begin{figure}[htb!]
  \begin{center}
    \epsfig{figure=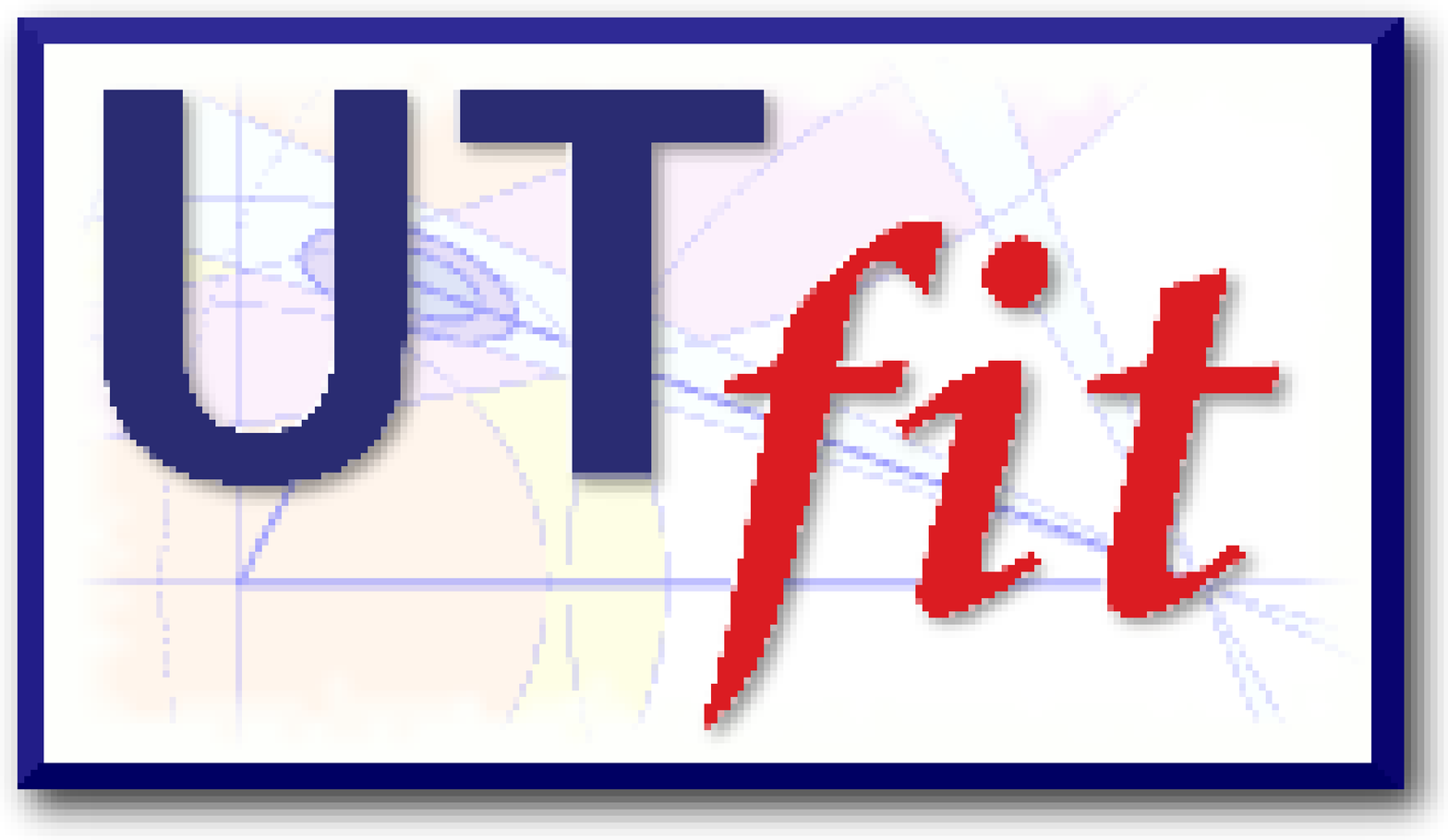,width=2.5cm}
  \end{center}
\end{figure}

\vspace*{-0.8cm}
%\vskip 1. cm
\begin{center}
  \Large{\textbf{UT}}\large{\textit{fit}}\large{~Collaboration :} \\
\end{center}
\begin{center}
  \begin{large}
  \textbf{M.~Bona$^{(a)}$, M.~Ciuchini$^{(b)}$, E.~Franco$^{(c)}$,
    V.~Lubicz$^{(b)}$, } \\
  \textbf{G. Martinelli$^{(c)}$, F. Parodi$^{(d)}$, M.
    Pierini$^{(e)}$, P.
    Roudeau$^{(f)}$, }\\
  \textbf{C. Schiavi$^{(d)}$, L.~Silvestrini$^{(c)}$, A.
    Stocchi$^{(f)}$ and V.~Vagnoni$^{(g)}$}
  \end{large}
\end{center}
\begin{center}
  \noindent 
  \begin{footnotesize}
    \noindent
    \textbf{$^{(a)}$   Laboratoire d'Annecy-le-Vieux de Physique des
      Particules,}\\ 
    \hspace*{0.5cm}{LAPP, IN2P3/CNRS, Universit\'e de Savoie, France}\\
    \textbf{$^{(b)}$ Dip. di Fisica, Universit{\`a} di Roma Tre
      and INFN,  Sez. di Roma III,}\\
    \hspace*{0.5cm}{Via della Vasca Navale 84, I-00146 Roma, Italy}\\
    \noindent
    \textbf{$^{(c)}$ Dip. di Fisica, Universit\`a di Roma ``La Sapienza'' and INFN, Sez. di Roma,}\\
    \hspace*{0.5cm}{Piazzale A. Moro 2, 00185 Roma, Italy}\\
    \noindent
    \textbf{$^{(d)}$ Dip. di Fisica, Universit\`a di Genova and INFN,}\\
    \hspace*{0.5cm}{Via Dodecaneso 33, 16146 Genova, Italy}\\
    \noindent
    \textbf{$^{(e)}$ Department of Physics, University of Wisconsin,}\\
    \hspace*{0.5cm}{Madison, WI 53706, USA}\\
    \noindent
    \textbf{$^{(f)}$ Laboratoire de l'Acc\'el\'erateur Lin\'eaire,}\\
    \hspace*{0.5cm}{IN2P3-CNRS et Universit\'e de Paris-Sud, BP 34,
      F-91898 Orsay Cedex, France}\\
    \textbf{$^{(g)}$INFN, Sez. di Bologna,}\\
    \hspace*{0.5cm}{Via Irnerio 46, I-40126 Bologna, Italy}\\
  \end{footnotesize}
\end{center}

\vspace*{0.5cm}

\begin{abstract}  
  The recent measurements of the $B^0_s$ meson mixing amplitude by CDF
  and of the leptonic branching fraction $BR(B \to \tau \nu_\tau)$ by
  Belle call for an upgraded analysis of the Unitarity Triangle in the
  Standard Model. Besides improving the previous constraints on the
  parameters of the CKM matrix, these new measurements, combined with
  the recent determinations of the angles $\alpha$, $\beta$ and
  $\gamma$ from non-leptonic decays, allow, in the Standard Model, a
  quite accurate extraction of the values of the hadronic matrix
  elements relevant for $K^0$-$\bar K^0$ and $B_{s,d}^0$-$\bar
  B_{s,d}^0$ mixing and of the leptonic decay constant $f_B$.  These
  values, obtained ``experimentally'', can then be compared with the
  theoretical predictions, mainly from lattice QCD.  In this paper we
  upgrade the UT fit, we determine from the data the kaon
  $B$-parameter $\hat B_K$, the $B^0$ mixing amplitude parameters
  $f_{B_{s}} \, \hat B^{1/2}_{B_{s}}$ and $\xi$, the decay constant
  $f_B$, and make a comparison of the obtained values with lattice
  predictions.  We also discuss the different determinations of
  $V_{ub}$ and show that current data do not favour the value measured
  in inclusive decays.
\end{abstract}

\newpage \pagestyle{plain}

\section{Introduction}
Lattice QCD (LQCD) played a relevant role in the history the Unitarity
Triangle (UT) fit since the very
beginning~\cite{prei1,preistoria,paganini,utfitseminal}, allowing
predictions of the value of $\sin 2\beta$ before the advent of direct
measurements by Babar and Belle~\cite{babarsin2b,bellesin2b}.  At the
time when the B factories had not started yet and inclusive
measurements of $\vub$ and $\vcb$ were rather rough, the ``classical''
UT analysis for the determination of $\rhobar$ and $\etabar$ relied on
the results of quenched lattice QCD simulations to relate the measured
exclusive semileptonic $B$ decays, the $B^0_{d}$--$\bar B^0_{d}$
mixing amplitude, the lower bound on $B^0_{s}$--$\bar B^0_{s}$
oscillations and CP violation in $K^0$--$\bar K^0$ mixing to the CKM
parameters. In spite of these caveat our prediction of $\sin 2 \beta$
in the years was quite stable, going from $\sin 2\beta= 0.65 \pm 0.12$
in 1995~\cite{prei1} to $\sin 2\beta= 0.698 \pm 0.066$ in
2000~\cite{utfitseminal}.

\begin{figure}[t]
\begin{center}
\includegraphics*[width=0.65\textwidth]{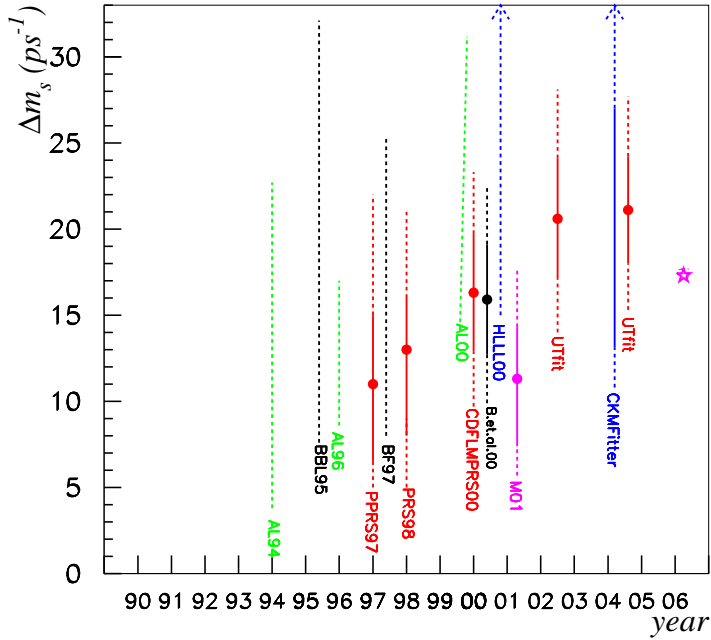}
\caption{\it Evolution of the ``indirect'' determination of $\Delta
  m_s$ over the years.  These determinations are given in
  \cite{utfitseminal,paganini,ref:allrhoeta,bargiotti,CKMfitterlatest,UTSM}.
  From left to right, they correspond to the following papers: AL94
  (Ali, London), BBL95 (Buchalla,Buras,Lautenbacher), AL96, PPRS97
  (Paganini, Parodi, Roudeau, Stocchi), BF97 (Buras,Fleischer), PRS98
  (Parodi,Roudeau,Stocchi), AL00, CDFLMPRS00 (Ciuchini et al.),
  B.et.al.00 (Bargiotti et al.), HLLL00
  (Hoecker,Laplace,Lacker,LeDiberder), M01 (Mele), UTFit (Bona et
  al.).  CKMFitter (J.Charles et al.).  The full (dotted) lines
  correspond to the 68$\%$(95$\%$) probability regions.  The star (for
  year '06) corresponds to the recent measured value by
  CDF~\cite{CDFDBS}. The error of the experimental measurement cannot
  be appreciated with this scale.}
\label{fig:storiadms}
\end{center}
\end{figure}

A similar situation is true for $\Delta m_s$, for which a first
precise indirect determination from the other constraints of the UT
fit was available since 1997 ($[6.5,15.0]$~ps$^{-1}$ at $68\%$
probability and $\Delta m_s<22$~ps$^{-1}$ at $95\%$
probability)~\cite{paganini}.  A compilation of the predictions for
$\Delta m_{s}$ by various collaborations as a function of time is
shown in Fig.~\ref{fig:storiadms}. As can be seen from this figure,
even in recent years, and despite the improved measurements, in some
approaches~\cite{bargiotti,CKMfitterlatest} the predicted range was
very large (or corresponds only to a lower bound~\cite{bargiotti}).
An upgraded version of our Standard Model ``prediction'' for $\Delta
m_s$, obtained from an overall UT fit which makes use of all the
latest input values and constraints, is given in the fifth column of
Tab.~\ref{tab:CKM_fit_today}: $\Delta m_s= (20.9 \pm 2.6)$~ps$^{-1}$.
This is the number and uncertainty to compare with the direct CDF
measurement given in eq.~(\ref{newexp}) below. Besides, in
Fig.~\ref{fig:pull1} we also show the compatibility plot for $\Delta
m_s$~\cite{UTSM}.
\begin{figure}[t]
\begin{center}
\includegraphics*[width=0.65\textwidth]{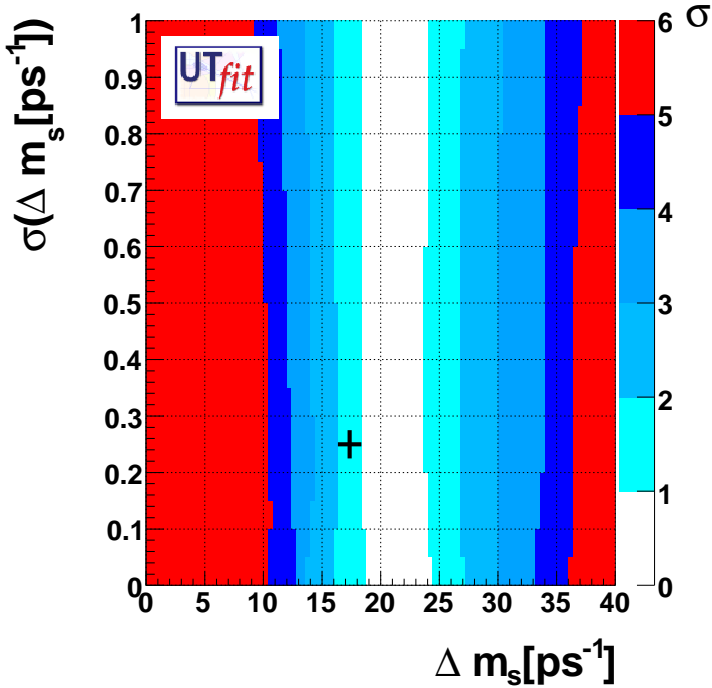}
\caption{\it Compatibility plot of the value of $\Delta m_s$ measured
  by CDF, $\Delta m_s= (17.33^{+0.42}_{-0.21} \,\,(\mathrm{stat}.)\,\,
  \pm 0.07\,\,(\mathrm{syst}.)) \,\, \mathrm{ps}^{-1}$ with the
  upgraded ``prediction'' from the other constraints of the Standard
  Model UT fit.}
\label{fig:pull1}
\end{center}
\end{figure}

More recently, we got much more information coming from the
determination of the UT angles, obtained by studying non-leptonic
decays: the angle $\alpha$ from $B\to \pi\pi$, $B\to \pi\rho$ and
$B\to \rho\rho$ decays~\cite{alpha}; the angle $\gamma$ from $B\to
D^{(*)}\, K^{(*)}$ decays~\cite{gamma}; $2\beta+\gamma$ from
time-dependent asymmetries in $B \to D^{(*)} \pi (\rho)$
decays~\cite{2bpg}; $\cos 2 \beta $ from $B^0_d \to J/\psi
K^{*0}_S$~\cite{cos2b}; $\beta$ from $B \to D^0\pi^0$~\cite{D0p0} and,
finally, $\sin 2 \beta$ from the ``golden mode'' $B^0_d \to J/\psi
K_S$~\cite{sin2b}.  In the following we will call the ensemble of
these measurements \utangles: they allow a determination of $\rhobar$
and $\etabar$ independently of the hadronic parameters computed on the
lattice. The precision in constraining $\rhobar$ and $\etabar$ from
the \utangles\ is by now comparable to that obtained from
lattice-related constraints, denoted as \utlattice.  The latter
include, besides the information coming from semileptonic decays,
namely $\vert V_{ub}\vert /\vert V_{cb}\vert$, the experimental
quantities $\epsilon_K$, $\Delta m_d$ and $\Delta m_s$.
 
The recent measurements of the neutral $B_s$ meson mixing amplitude by
the CDF Collaboration~\cite{CDFDBS}, and of the leptonic branching
fraction $BR(B \to \tau \nu_\tau)$ by the Belle Collaboration~\cite{BellefB}
\begin{eqnarray} 
  \Delta m_s  & =  & (17.33^{+0.42}_{-0.21} \,\,(\mathrm{stat}.)\,\, \pm
  0.07\,\,(\mathrm{syst}.)) \,\, \mathrm{ps}^{-1}\quad  {\rm CDF} \nonumber  \\
  BR(B \to \tau \nu_\tau) & =& (1.06 ^{+0.34}_{-0.28} \,\,(\mathrm{stat}.)\,\, ^{
    +0.18}_{-0.16}\,\,(\mathrm{syst}.)) \times 10^{-4}\quad {\rm Belle}
   \, ,  \label{newexp}
\end{eqnarray} 
and the additional bounds given respectively by the D0~\cite{D0DBS}
and BaBar~\cite{BabarfB} Collaborations, provide further information
for the analysis of the Unitarity Triangle in the Standard Model.  In
this paper, besides improving the determination of the constraints on
the parameters of the CKM matrix via the standard UT analysis, we show
that the new measurements allow a quite accurate extraction of the
values of the hadronic matrix elements relevant for $K^0$-$\bar K^0$
and $B_{s,d}^0$-$\bar B_{s,d}^0$ mixing and of the leptonic decay
constant $f_B$.  Assuming that there is no contribution from New
Physics, we determine these hadronic quantities from the experimental
data and compare them with recent lattice
calculations~\cite{Dawson:2005za,hashimoto}. We also discuss the
different determinations of $V_{ub}$ and show that there is an
indication that the value measured in inclusive decays is not favoured
by the data.
 
\section{Upgraded \utfit\ Analysis} 
In this section we give the results of the upgraded analysis which
includes the new measurement of $\Delta m_s$ by the CDF Collaboration.
This result improves the determination of $\Delta m_s$ by LEP, SLD and
previous TeVatron analyses~\cite{D0DBS,HFAG}. Given the uncertainty on
the theoretical value of $f_B$ and the still relatively large error in
the experimental measurement, the effect of $BR(B \to \tau \nu_\tau)$
on the analysis is negligible at this stage.  Indeed by taking from
the lattice $f_B=(189 \pm 27 )$~MeV~\cite{hashimoto}, one gets $\vert
V_{ub}\vert=(41\pm9) \times 10^{-4}$ with an error much larger than
the uncertainty of determinations from exclusive or inclusive
semileptonic decays.

In Tab.~\ref{tab:input} we give the value of the upgraded input
parameters.  In some cases the same quantities, e.g. $\sin 2\beta$,
also appear, with a different central value and uncertainty, in
Tab.~\ref {tab:CKM_fit_today}, where we give the output results of the
UT fit. The reason is that the final output values of Tab.~\ref
{tab:CKM_fit_today} are obtained by combining all the available
information on a given quantity~\cite{paganini,utfitseminal,UTSM}: in
the case of $\sin 2\beta$, for example, the information coming from
the \utangles \ and \utlattice \ measurements.

\begin{table*}[htbp!]
{\footnotesize
\begin{center}
\begin{tabular}{@{}llll}
\hline\hline
         Parameter                          &  Value                            
     & Gaussian ($\sigma$)      &   Uniform             \\
                                            &                                   
     &                          & (half-width)          \\ \hline\hline
         $\lambda$                          &  0.2258                           
     &  0.0014                  &    -                  \\ \hline
$\left |V_{cb} \right |$(excl.)             & $ 41.4 \times 10^{-3}$            
     & $2.1 \times 10^{-3}$     & -                     \\
$\left |V_{cb} \right |$(incl.)             & $ 41.6 \times 10^{-3}$            
     & $0.7 \times 10^{-3}$     & $0.6 \times 10^{-3}$  \\ 
$\left |V_{ub} \right |$(excl.)             & $ 38.0  \times 10^{-4}$           
     & $2.7 \times 10^{-4}$     & $4.7 \times 10^{-4}$  \\  
$\left |V_{ub} \right |$(incl.)        & $ 44.5  \times 10^{-4}$           
     & $2.0 \times 10^{-4}$     &        $2.6 \times 10^{-4}$            \\ \hline
$\Delta m_d$                                & $0.502~\mbox{ps}^{-1}$            
     & $0.006~\mbox{ps}^{-1}$   &        -              \\
$\Delta m_s$                                & $ 17.35$  ps$^{-1}$  &   $ ^{+0.42}_{-0.21}\pm0.07$  ps$^{-1}$
     & -   \\ \hline
$\fbssqbs$                                  & $262$ MeV                        
     & $35$ MeV                 &          -            \\
$\xi=\frac{\fbssqbs}{\fbdsqbd}$             & 1.23                              
     & 0.06                     &  -            \\\hline
$\hat B_K$                                  & 0.79                              
     & 0.04                     &     0.08              \\
$\epsilonk$                                 & $2.280 \times 10^{-3}$            
     & $0.013 \times 10^{-3}$   &          -            \\
$f_K$                                       & 0.159 GeV                         
     & \multicolumn{2}{c}{fixed}                        \\
$\Delta m_K$                                & 0.5301 $\times 10^{-2}
~\mbox{ps}^{-1}$ & \multicolumn{2}{c}{fixed}                        \\ \hline
$\snb$                                      &  0.687              
     &  0.032                   &          -            \\ \hline
$\overline m_t$                                       & $168.5$ GeV
     & $4.1$ GeV          &          -            \\
$\overline m_b$                                       & 4.21 GeV
     & 0.08 GeV           &          -            \\
$\overline m_c$                                       & 1.3 GeV
     & 0.1 GeV            &          -            \\
$\alpha_s(M_Z)$                                  & 0.119                             
     & 0.003                    &          -            \\
$G_F $                                      & 1.16639 $\times 10^{-5} \GeV^{-2}$
     & \multicolumn{2}{c}{fixed}                        \\
$ m_{W}$                                    & 80.425 GeV
     & \multicolumn{2}{c}{fixed}                        \\
$ m_{B^0_d}$                                & 5.279 GeV
     & \multicolumn{2}{c}{fixed}                        \\
$ m_{B^0_s}$                                & 5.375 GeV
     & \multicolumn{2}{c}{fixed}                        \\
$ m_K^0$                                   & 0.497648 GeV
     & \multicolumn{2}{c}{fixed}                        \\ \hline\hline
\end{tabular} 
\end{center}
}
\caption {\it {Values of the relevant input quantities used in the UT fit.
    The Gaussian and the flat contributions to the
    uncertainty are given in the third and fourth columns 
    respectively (for details on the statistical treatment see~\cite{utfitseminal}). 
  }}
\label{tab:input} 
\end{table*}

\begin{figure}[htb!]
\begin{center}
\includegraphics*[width=0.65\textwidth]{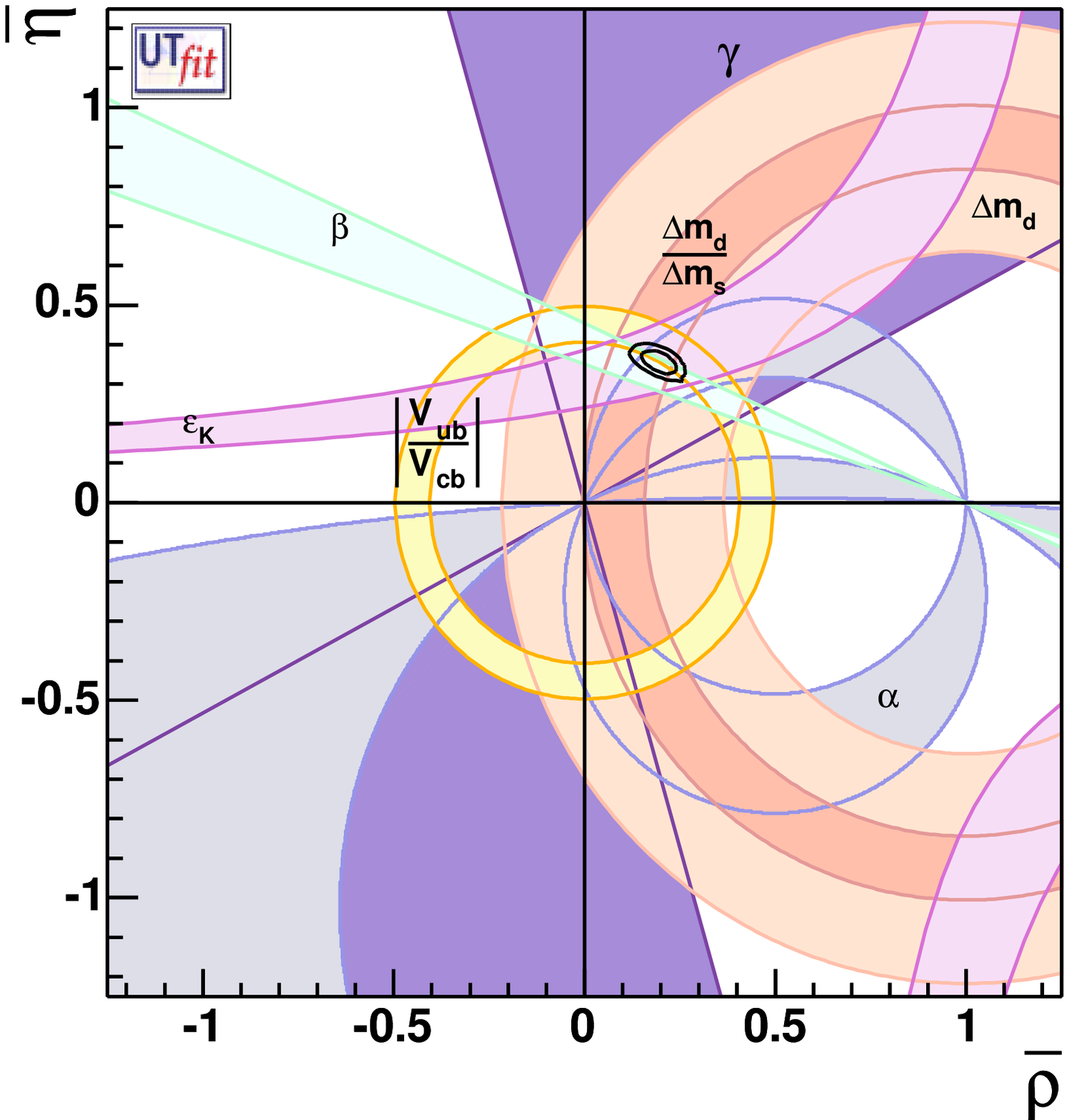}
\caption{%
  \textit{Determination of $\rhobar$ and $\etabar$ from constraints on
    $\left | V_{ub} \right |/\left | V_{cb} \right |$, $\Delta {m_d}$,
    $\Delta {m_s}$, $\epsilonk$, $\beta$, $\gamma$, and $\alpha$.
    $68\%$ and $95\%$ total probability contours are shown, together
    with $95\%$ probability regions from the individual constraints.}}
\label{fig:CKM_fit_today}
\end{center}
\end{figure}

In Fig.~\ref{fig:CKM_fit_today} we show the results of the new fit
which includes all constraints: $\left | V_{ub} \right |/\left |
  V_{cb} \right |$, $\Delta {m_d}$, $\Delta {m_s}$, $\epsilonk$,
$\alpha$, $\beta$, and $\gamma$.  In addition in
Tab.~\ref{tab:CKM_fit_today} we present for comparison the values and
uncertainties of the relevant quantities for the two cases, \utangles\
and \utlattice, whereas in the column labelled as ``All'' we give the
results of the analysis including all constraints.~\footnote{For
  further details on the UT analysis of the UTfit Collaboration see
  refs.~\cite{utfitseminal,UTSM,sitoweb}; for the results of the
  CKMfitter collaboration see~\cite{CKMfitterlatest,charles}.}

\begin{table*}[h]
{\footnotesize
\begin{center}
\begin{tabular}{@{}cccccccc}
\hline\hline  
    Parameter        &     \utangles     &  \utlattice   &  All            & All[no $\Delta m_s$]     & All[$V_{ub}$-excl]  &All[$V_{ub}$-incl] \\\hline
$\overline {\rho}$   & $0.204\pm 0.055$ &$0.197\pm0.035$ &$0.197\pm 0.031$ &$0.228\pm 0.034$ &$0.167\pm 0.031$ &$0.197\pm 0.032$ \\\hline
$\overline {\eta}$   & $0.317\pm 0.025$ &$0.389\pm0.025$ &$0.351\pm 0.020$ &$0.336\pm 0.021$ &$0.334\pm 0.018$ &$0.351\pm 0.020$ \\\hline
$\alpha [^{\circ}]$  & $100  \pm 8~$    &$90.8 \pm4.9$   &$95.5 \pm 4.8$   &$99.5  \pm 4.5~$ &$94.4 \pm 4.6$   &$95.5 \pm 4.9$   \\\hline
$\beta [^{\circ}]$   & $21.8 \pm 1.3~$  &$25.8 \pm1.4$   &$23.6\pm 1.0$    &$21.8 \pm 1.3~$  &$21.8 \pm 1.1$   &$23.5 \pm 1.0$   \\\hline
$\sin 2 \beta$       & $0.687\pm 0.032$ &$0.784\pm0.032$ &$0.733\pm 0.024$ &$0.730\pm 0.023$ &$0.689\pm 0.028$ &$0.734\pm 0.024$ \\\hline
$\sin 2 \beta_s$     & $0.034\pm 0.003$ &$0.041\pm0.003$ &$0.037\pm 0.002$ &$0.036\pm 0.002$ &$0.036\pm 0.002$ &$0.038\pm 0.002$ \\\hline
$\gamma[^{\circ}$]   & $57.4 \pm8.4$    &$63.0 \pm4.8~$  &$60.6 \pm 4.7$   &$55.8 \pm 5.2$   &$63.5 \pm 4.6$   &$60.7 \pm 4.8$   \\\hline
$\rm{Im} {\lambda}_t$[$10^{-5}$] 
                     &  $12.6\pm1.1~$   &$15.3\pm0.9~$   &$14.1 \pm0.7$    &$13.7 \pm0.8$    &$13.3 \pm0.7$    &$14.2 \pm0.08$   \\\hline
$\Delta m_s$[ps$^{-1}$]  
                     & $20    \pm5~$    &$17.4\pm0.3~$   &$17.5 \pm 0.3$  &$20.9 \pm 2.6$     &$17.4 \pm 0.3$   &$17.4 \pm 0.3$   \\\hline
$V_{ub}[10^{-3}]$    & $3.67\pm0.24$    &$4.18\pm0.20$   &$3.91 \pm0.14$  &$3.96 \pm0.14$    &$3.60 \pm0.17$   &$3.92 \pm0.16$   \\\hline
$V_{cb}[10^{-2}]$    & $4.15\pm0.07$    &$4.12\pm0.07$   &$4.17 \pm0.06$  &$4.19 \pm0.06$    &$4.15 \pm0.06$   &$4.17 \pm0.06$   \\\hline
$V_{td}[10^{-3}]$    & $8.03\pm0.57$    &$8.30\pm0.31$   &$8.26 \pm0.31$  &$7.97 \pm0.34$    &$8.43 \pm0.28$   &$8.26 \pm0.32$   \\\hline
$|V_{td}/V_{ts}|$    & $0.197\pm0.015$  &$0.205\pm0.009$ &$0.201\pm0.008$ &$0.192\pm0.009$   &$0.206\pm0.007$  &$0.201\pm0.008$  \\\hline
$R_b$                & $0.382\pm0.024$  &$0.438\pm0.023$ &$0.404\pm0.015$ &$0.408\pm0.015$   &$0.374\pm0.018$  &$0.404\pm0.016$  \\\hline
$R_t$                & $0.856\pm0.058$  &$0.891\pm0.036$ &$0.875\pm0.034$ &$0.841\pm0.037$   &$0.897\pm0.031$  &$0.875\pm0.034$  \\
\hline\hline
\end{tabular} 
\end{center}}
\caption{\textit{Comparison of determinations of UT parameters
    from the constraints on the angles $\alpha$, $\beta$, and $\gamma$ 
    (\utangles) and from lattice-dependent quantities $\vert
    V_{ub}/V_{cb}\vert$, $\Delta m_d$, 
    $\Delta m_s$, and $\epsilon_K$ (\utlattice). We also show the results
    obtained by using all the constraints together} (All), \textit{all the
    constraints except  $\Delta m_s$} (All[no $\Delta m_s$]), \textit{all the
    constraints except the inclusive $\vert V_{ub}\vert$}
    (All[$V_{ub}$-excl]) \textit{and 
    all the constraints except the exclusive $\vert V_{ub}\vert$}
    (All[$V_{ub}$-incl]). \textit{For the definition of $R_b$ and $R_t$
    see for example ref.~\cite{Battaglia:2003in},  
    for the definition of $\sin 2\beta_s$ see ref.~\cite{Nir:2005js}.}} 
\label{tab:CKM_fit_today}
\end{table*}

 Several observations are important at this point:
 \begin{itemize}  
 \item The recent measurement of $\Delta m_s$ reduces the
   uncertainties, although not in a dramatic way.
 \item If we compare Tab.~\ref{tab:input} and
   Tab.~\ref{tab:CKM_fit_today} with the corresponding ones of our
   previous published UT analysis~\cite{UTSM}, we note that the
   directly measured value of $\sin 2\beta$ has decreased from $\sin
   2\beta=0.726(37)$ (old) to $\sin 2\beta=0.687(32)$ (new). 
As a consequence, 
   the overlap between the regions of the $\bar
   \rho$-$\bar \eta$ plane, selected by  the
   \utangles\ with respect to the region selected by the \utlattice,
is reduced.
   This is shown in Fig.~\ref{fig:latticevsangles}  where we
   superimpose the region selected by the \utangles\ to the 68\% and 
   95\% probability contours coming from the \utlattice\ fit. A similar
   figure with 2004 data would have given a much better agreement.
   Besides the fact that the measurements are now more precise, the
   worse agreement is due to i) the lower value of $\sin 2\beta$ and
   ii)  an important  reduction of the quoted uncertainty of the
   inclusive $\vert V_{ub}\vert$. 
  \begin{figure}[htb!]
\begin{center}
\includegraphics*[width=0.65\textwidth]{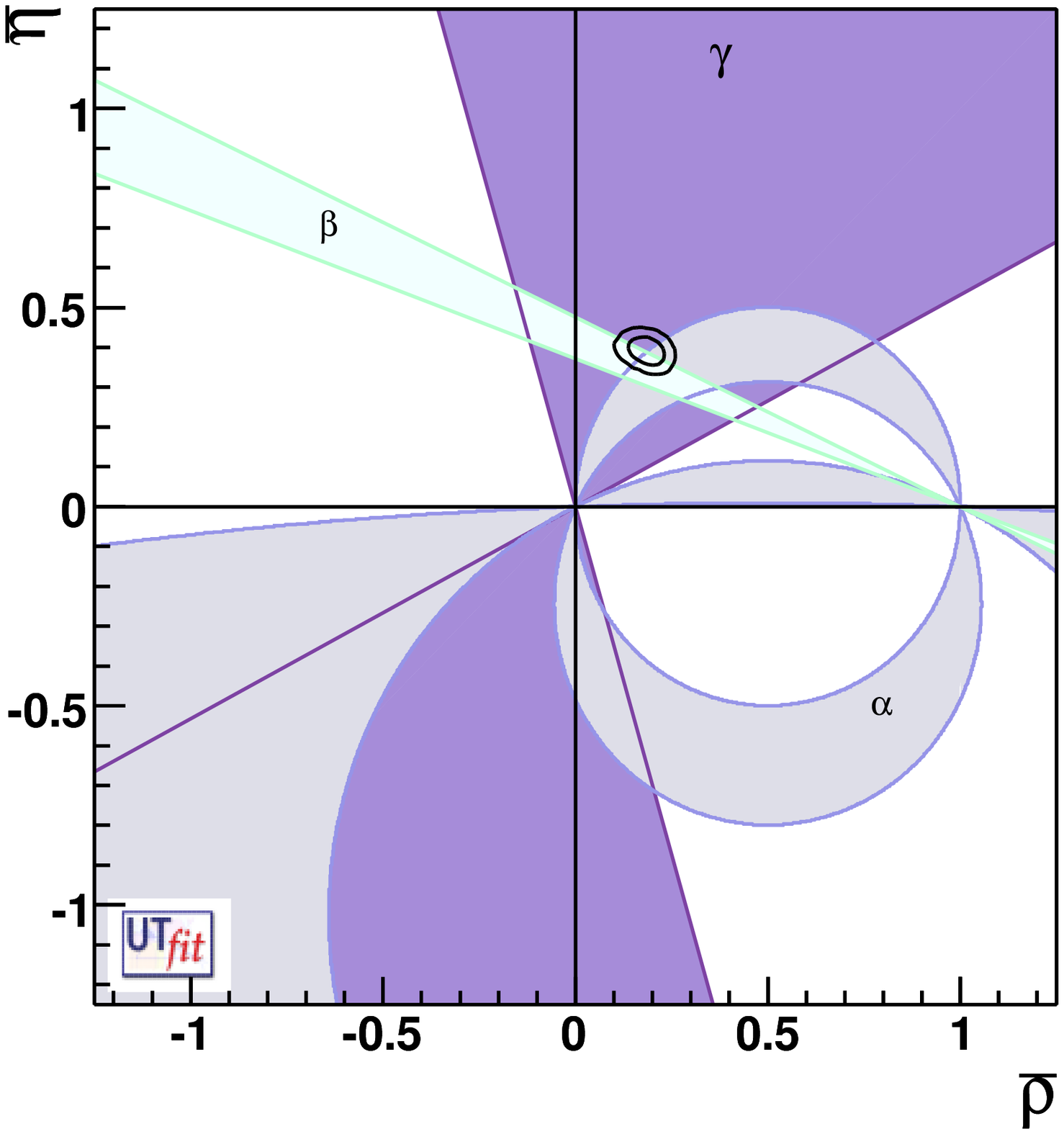}
\caption{%
  \textit{Determination of $\rhobar$ and $\etabar$ from constraints on
    $\left | V_{ub} \right |/\left | V_{cb} \right |$, $\Delta {m_d}$,
    $\Delta {m_s}$ and $\epsilonk$ ($68\%$ and $95\%$ total
    probability contours), compared to the $95\%$ probability regions
    of the individual constraints on $\beta$, $\gamma$, and
    $\alpha$.}}
\label{fig:latticevsangles}
\end{center}
\end{figure}
 \item The difference between the results with \utangles\ and \utlattice\ 
   is also demonstrated by a comparison of the experimental value,
   $\sin 2\beta=0.687(32)$,   with the value obtained by using only the 
   \utlattice\ measurements,   $\sin 2\beta_{\rm UTlattice}=0.784(32)$.
\item $\bar \eta$ is also  an instructive quantity to visualize the
   important difference between the \utangles\ result, $\bar
    \eta_{\rm UTangles} = 0.317 \pm 0.025$ and the \utlattice\ case,
    $\bar \eta_{\rm UTlattice}=0.389 \pm 0.025$.

  \item In order to understand where these differences come from, we
    have studied the correlation between the value of $\sin
    2\beta_{\rm UTlattice}$ and $\vert V_{ub}\vert$ with the following
    results: if we use only the exclusive value of $\vert
    V_{ub}\vert$, we get $\sin 2\beta_{\rm
      UTlattice-excl.}=0.704(55)$, much closer to $\sin 2\beta_{\rm
      UTangles}=0.687(32)$ whereas if we use only the inclusive value
    of $\vert V_{ub}\vert$ we obtain $\sin 2\beta_{\rm
      UTlattice-incl.}=0.804(37)$. This implies that there is a strong
    correlation between $\vert V_{ub}\vert$ and $\sin 2 \beta_{\rm
      UTlattice}$. This is true also for $\bar \eta$ as shown by a
    comparison between $\bar \eta_{\rm UTlattice-excl.}=0.349\pm0.032$
    and $\bar \eta_{\rm UTlattice-incl.}=0.400 \pm 0.028$.  To
    investigate further this point we performed the complete UT fit
    either using only the exclusive value of $\vert V_{ub}\vert$
    (All[$V_{ub}$-excl]) or only the inclusive one
    (All[$V_{ub}$-incl]).  In the left (right) plot of
    Fig.~\ref{fig:pull_vubexcl}, we give for All[$V_{ub}$-excl]
    (All[$V_{ub}$-incl]) the compatibility plot~\cite{UTSM} for the
    inclusive (exclusive) determination of $\vert V_{ub}\vert$.  We
    conclude that the inclusive value of $\vert V_{ub}\vert$ is not in
    agreement with the determination of $\vert V_{ub}\vert $ from all
    other constraints, at the 2.5$\sigma$ level.
\begin{figure}[htb!]
\begin{center}
\includegraphics*[width=0.45\textwidth]{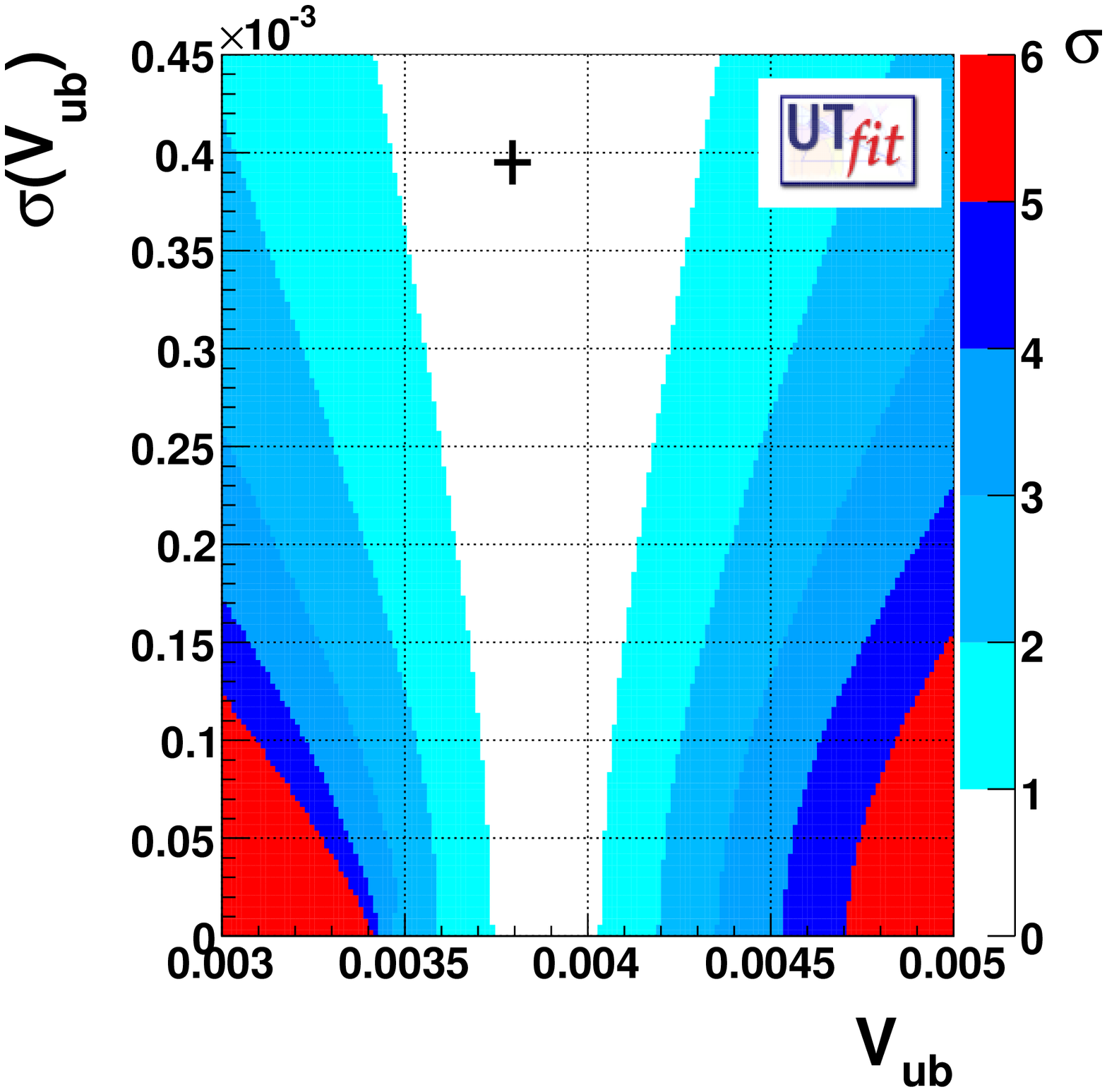}
\includegraphics*[width=0.45\textwidth]{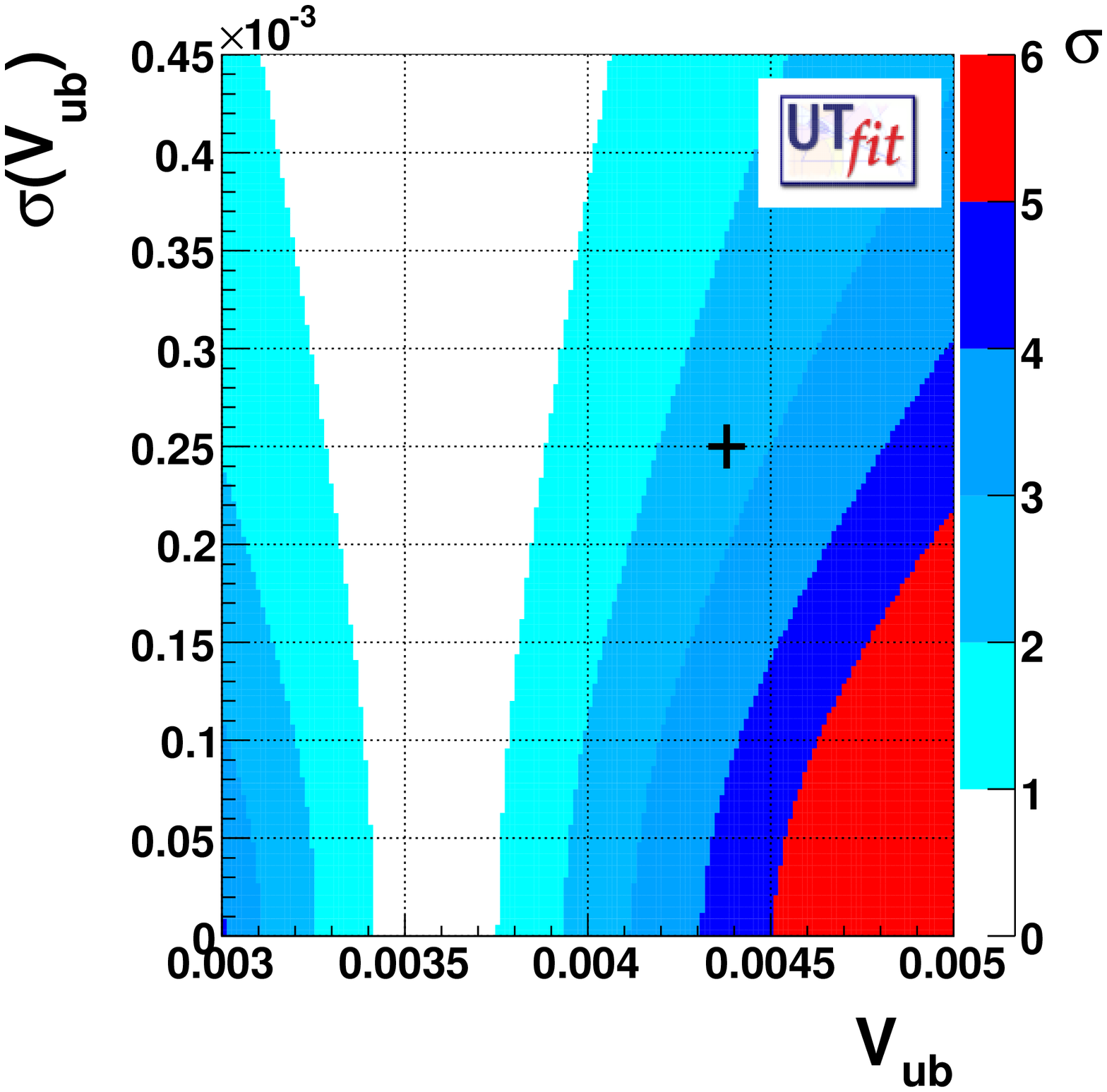}
\caption{%
  \textit{Left: Compatibility plot between the direct determination of
    $\vert V_{ub}\vert$ from exclusive analysis and the rest of the
    fit (including the constraint on $\vert V_{ub}\vert$ from
    inclusive analysis).  Right:Compatibility plot between the direct
    determination of $\vert V_{ub}\vert$ from inclusive analysis and
    the rest of the fit (including the constraint on $\vert
    V_{ub}\vert$ from exclusive analysis).}}
\label{fig:pull_vubexcl}
\end{center}
\end{figure}

\item In order to investigate whether the problem originates from a
  tension between the experimental value of $\sin 2 \beta$ and $\vert
  V_{ub}\vert $, we also present the compatibility plot for $\sin 2
  \beta$ including all other measurements (left plot of
  Fig.~\ref{fig:sin2bpull}) or all other measurements except $\vert
  V_{ub} \vert$ (right plot of Fig.~\ref{fig:sin2bpull}).  We conclude
  that rather than a problem between $\sin 2 \beta$ and $\vert
  V_{ub}\vert $, the tension arises between $\vert V_{ub}\vert $ and
  several quantities entering the UT fit.  A larger value of $\sin 2
  \beta$ would only soften the problem.

\begin{figure}[htb!]
\begin{center}
\includegraphics*[width=0.45\textwidth]{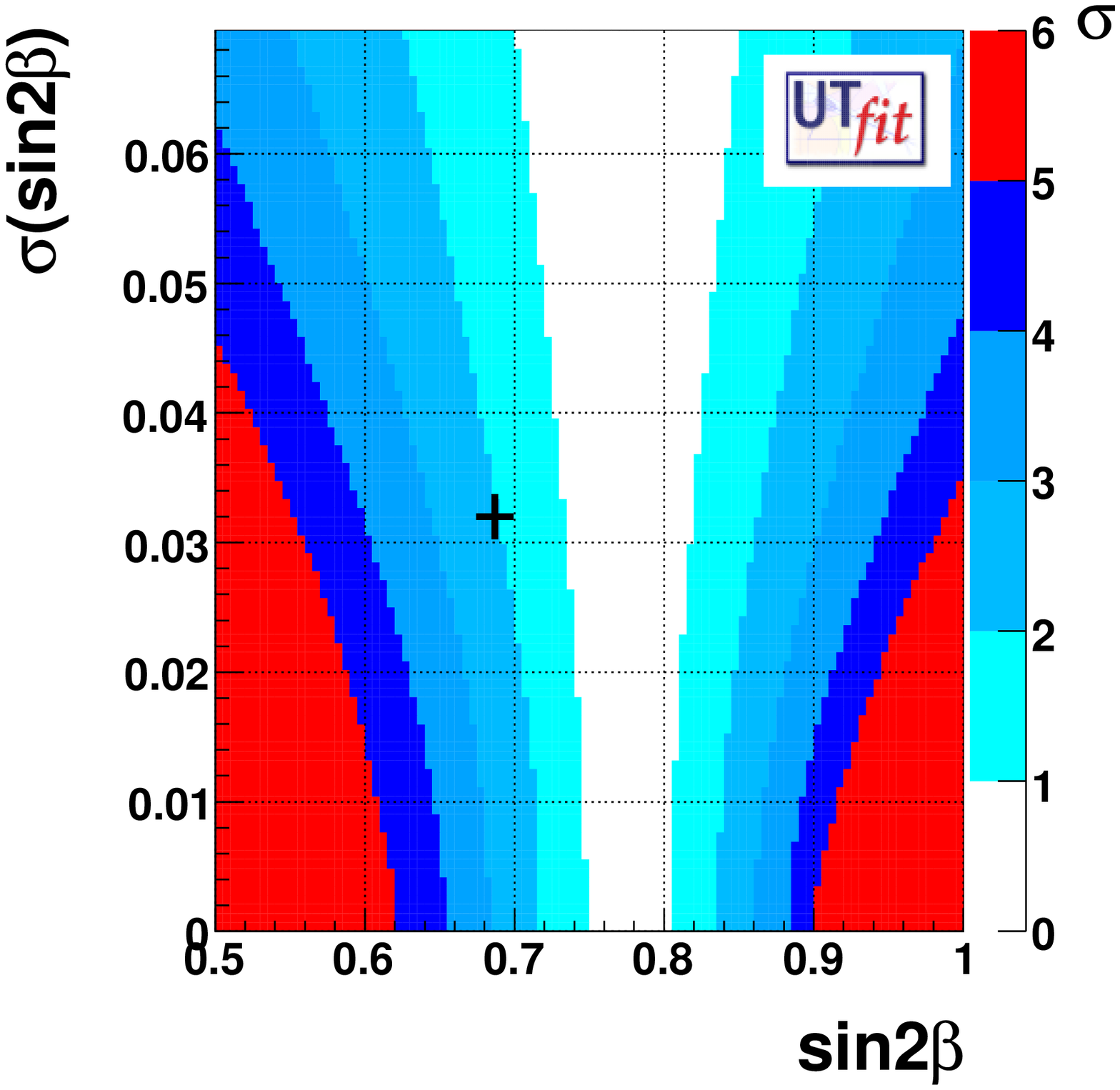}
\includegraphics*[width=0.45\textwidth]{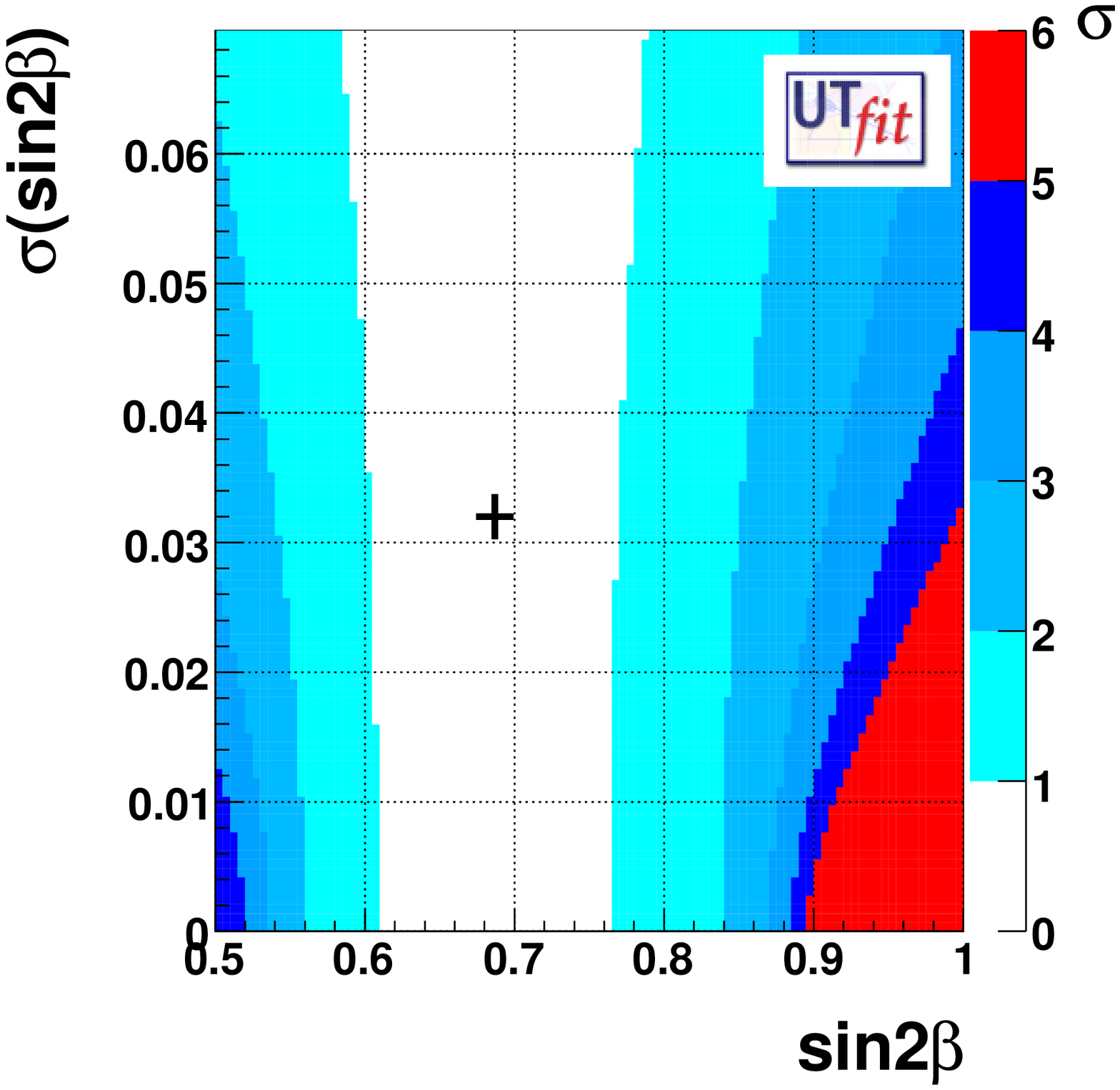}
\caption{%
  \textit{ Compatibility plot of the experimental value of
    $\sin2\beta$ (cross) and the prediction from the fit done with all
    the other information, using (left) or ignoring (right) the
    constraint from $\vert V_{ub} \vert$.\label{fig:sin2bpull}}}
\end{center}
\end{figure}

\item It is worth recalling that the value of $\vert V_{ub}\vert $
  that is extracted from the experiments also relies on non
  perturbative hadronic quantities (the semileptonic form factors
  $f^+(q^2)$, $V(q^2)$, $A_{1,2}(q^2)$ for exclusive $B \to \pi$ and
  $B \to \rho$ decays and the parameters $\bar \Lambda$, $\lambda_1$
  and $\lambda_2$ for inclusive semileptonic decays).  The systematic
  difference between the exclusive and inclusive determination of
  $\vert V_{ub}\vert $ (the inclusive values are always larger than
  the exclusive ones) might be explained by the uncertainties of the
  theoretical approaches. Our analysis suggests that, although all the
  results are still compatible, there could be some problem with the
  theoretical calculations, and/or with the estimate of the
  uncertainties, of inclusive $b \to u $ semileptonic decays.  On the
  other hand, an effort should be made to increase the precision on
  the form factor of $B \to \pi$ and $B\to \rho$, providing all of
  them in the unquenched case, with low light quark masses and
  studying the continuum limit of the relevant form factors.  Note
  that this tension among exclusive and inclusive calculations is a
  peculiarity of $\vert V_{ub} \vert$, since the inclusive and
  exclusive determinations of $\vert V_{cb}\vert $ are in much better
  agreement.
\item Not having used $BR(B \to \tau \nu_\tau)$ as an input in the
  analysis, we can indirectly determine its value as an output of our
  fit. This is obtained starting from the \utangles\ determination of
  $\bar\rho$ and $\bar\eta$, combined with the experimental
  determination of $\vert V_{ub}\vert$ and $\vert V_{cb}\vert$, adding
  the experimental measurement of $\Delta m_d$ and $\Delta m_s$ to
  determine $f_B\sqrt{B_{Bd}}$, and using the lattice value of $\hat
  B_{Bd}$, $\hat B_{Bd}=1.28 \pm 0.05 \pm 0.09$~\cite{hashimoto} to
  obtain $f_B$ from it.  In this way, the prediction is obtained
  without using the value of $f_B$ taken from lattice calculations,
  which has a larger relative uncertainty than $\hat B_{Bd}$.  In this
  way, we obtain the following values:
   \begin{eqnarray}
     \label{eq:btaunupred}
     BR(B \to \tau \nu_\tau)_\mathrm{All} &=& (1.41 \pm 0.33) \times10^{-4}\,, \\
     BR(B \to \tau \nu_\tau)_{V_{ub}-\mathrm{incl}} &=& (1.53 \pm 0.41)
     \times10^{-4} \,,
     \nonumber\\
     BR(B \to \tau \nu_\tau)_{V_{ub}-\mathrm{excl}} &=& (1.02\pm 0.22)
     \times10^{-4}\,. 
     \nonumber
   \end{eqnarray}

Although all the predictions above are compatible within the errors,
a comparison of the  values given in eq.~(\ref{eq:btaunupred}) gives the 
measure of the correlation of this prediction with $\vert V_{ub}\vert$ 
in the overall UT fit, since all other input quantities are the same.

For comparison, with $f_B=(189 \pm 27)$~MeV and $\vert V_{ub}\vert
=(4.2 \pm 0.3)\times 10^{-3}$, one would obtain $BR(B \to \tau
\nu_\tau) =(1.17 \pm 0.50)\times 10^{-4}$.  Note that also in this
case a better agreement between the prediction and the experimental
world average ($BR(B \to \tau \nu_\tau) = (1.08 \pm 0.24)\times
10^{-4}$, combining Belle~\cite{BellefB} and BaBar~\cite{BabarfB}) is
found when the exclusive value of $\vert V_{ub}\vert$, or the value
from UTangles, is used.  The p.d.f. for this quantity is given in
Fig.~\ref{fig:btaunu}.

It is important to improve the predictive power on this quantity and
to clarify the situation of the $\vert V_{ub}\vert$ input, since a
possible future discrepancy between the value of the experimental
measurement and the theoretical prediction could signal effects of new
physics from extra Higgs particles~\cite{gino}.
\item Another possibility is to predict $\Delta m_s$ without using the
  experimental value. 
In order to display also in this case the correlation with the value of $\vert
  V_{ub}\vert$, we consider several possibilities for $\vert V_{ub}\vert$:
   \begin{eqnarray}
     \label{eq:dmsprediction}
     \Delta m_s~(\mathrm{All})&=& (20.9 \pm 2.6) \mathrm{ps}^{-1}\,,\\
     \Delta m_s~(V_{ub}-\mathrm{excl})&=& (19.4 \pm 2.5) \mathrm{ps}^{-1}\,, \nonumber\\
     \Delta m_s~(V_{ub}-\mathrm{incl})&=& (21.7 \pm 2.8) \mathrm{ps}^{-1}\,. \nonumber
   \end{eqnarray}
\end{itemize}
\begin{figure}[htb!]
\begin{center}
\includegraphics*[width=0.65\textwidth]{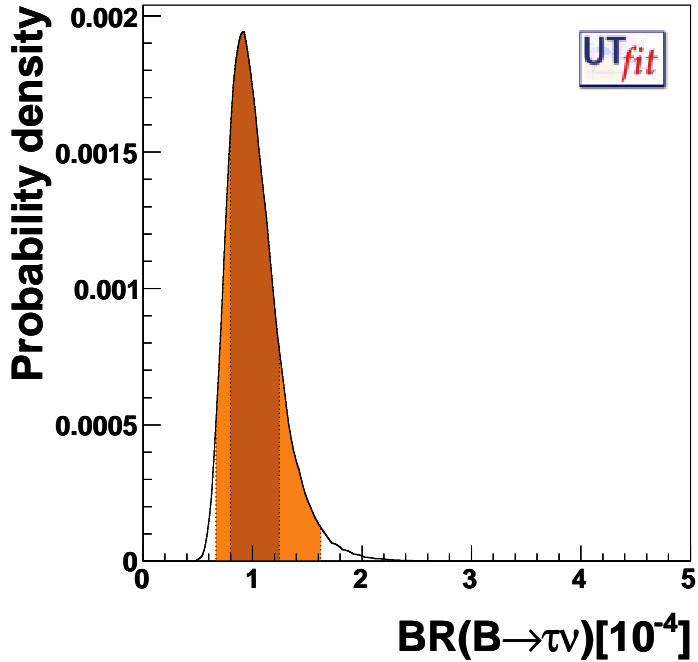}
\caption{%
  \textit{Determination of $BR(B \to \tau \nu_\tau)$ using the
    constraint from $\alpha$, $\beta$, $\gamma$, and $\vert
    V_{ub}/V_{cb} \vert$ to determine $\bar\rho$ and $\bar\eta$,
    $\Delta m_s$, and $\Delta m_d$ to fix the lattice parameters
    $\fbssqbs$ and $\xi$, and using $\hat B_{B_d}$ from lattice QCD.
    Only the exclusive determination of $\vert V_{ub}\vert$ is used in
    this case.}}
\label{fig:btaunu}
\end{center}
\end{figure}

\section{Constraints on Lattice Parameters}
Assuming the validity of the Standard Model, the constraints in the
$\bar \rho$-$\bar \eta$ plane from \utangles\ and semileptonic $B$
decay measurements, combined with the experimental values of $\Delta
m_d$, $\Delta m_s$ and $\epsilon_K$, allow the ``experimental''
determination of several hadronic quantities which were previously
taken from lattice QCD calculations. This approach has two important
advantages. The first one is that we have the possibility of making a
full UT analysis without relying at all on theoretical calculations of
hadronic matrix elements, for which there was a long debate about the
treatment of values and error distributions. The second advantage is
that we can extract from the combined experimental measurements the
value of $\hat B_K$ and of the $B^0$ mixing amplitudes $f_{B_{s,d}} \,
\hat B^{1/2}_{B_{s,d}}$ (or equivalently $f_{B_{s}} \, \hat
B^{1/2}_{B_{s}}$ and $\xi$) and compare them to the theoretical
predictions.

\begin{figure}[t]
  \centering
  \includegraphics[width=0.8\textwidth]{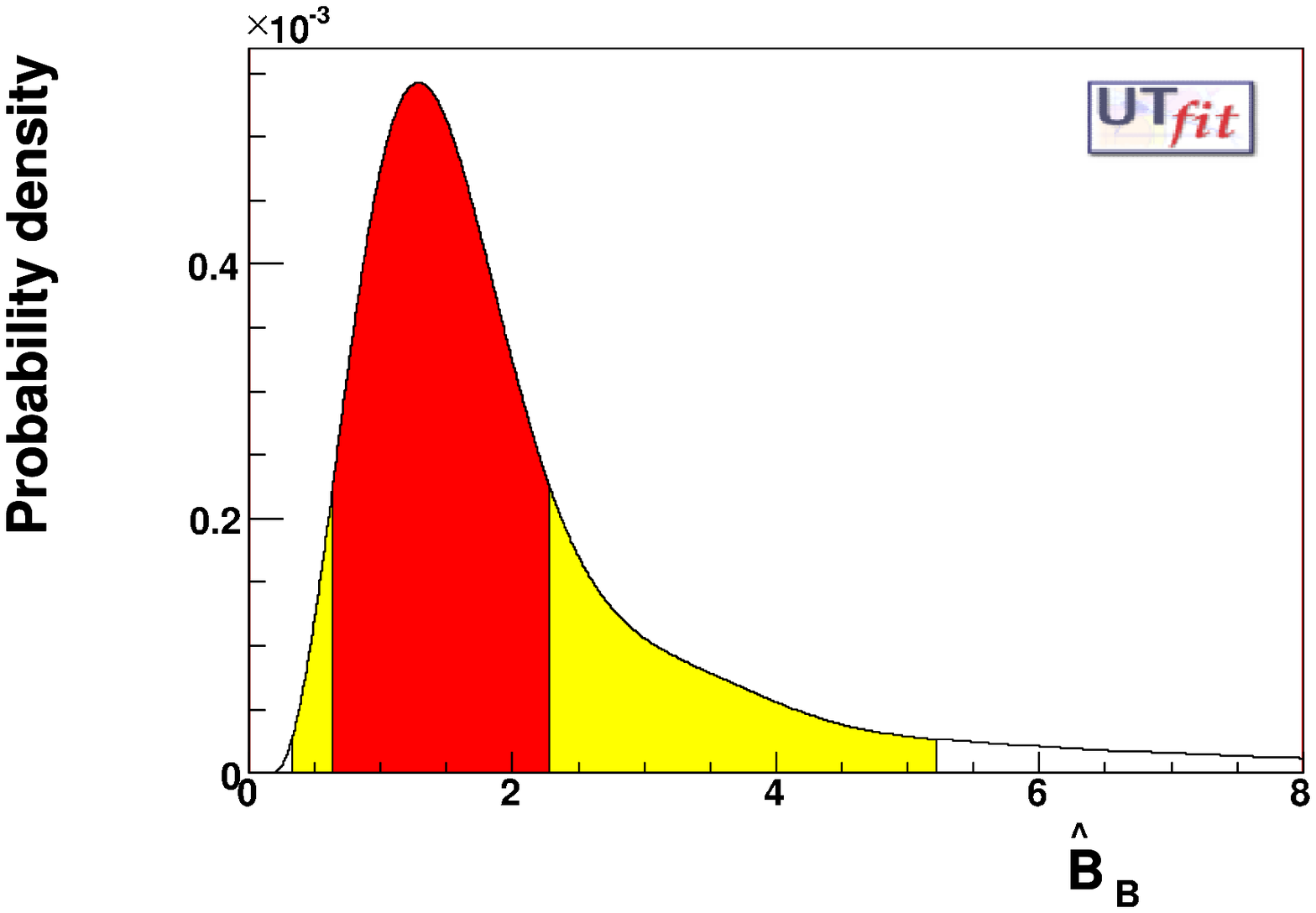}
  \caption{\it P.d.f. for $\hat B_{B_{d}}$ extracted from the UT analysis using $BR(B \to \tau \nu_\tau)$ to determine $f_B$.}
  \label{fig:Bbhat}
\end{figure}

Besides $\hat B_K$, $f_{B_{s}} \, \hat B^{1/2}_{B_{s}}$ and $\xi$, the
measurement of $BR(B \to \tau \nu_\tau)$ also allows a test of the
theory for the leptonic decay constant $f_B$, which is one of the
ingredients used by lattice calculations to predict the mixing matrix
element (proportional to $f^2_{B} \, \hat B_{B}$).  Finally by
combining the measurement of $BR(B \to \tau \nu_\tau)$ with $\Delta
m_d$ and the knowledge of the angles, we can extract the value of
$\hat B_{B_{d}}$ and compare with lattice predictions. In this case,
because of the experimental error on $BR(B \to \tau \nu_\tau)$, we
obtain a p.d.f. for $\hat B_{B_{d}}$ with a long tail (see
Fig.~\ref{fig:Bbhat}), corresponding to $\hat B_{B_{d}} = 2.1 \pm
1.0$,\footnote{This result is obtain using the median, which is
  appropriate given the long tail of the distribution. Using instead
  the mean we would obtain $\hat B_{B_{d}} = 1.5 \pm 0.8$.} which then
is not yet competitive with the lattice prediction, $\hat B_{Bd}=1.28
\pm 0.05 \pm 0.09$~\cite{hashimoto}.  Since the results depend on the
input value for $\vert V_{ub}\vert$, we consider two cases: all the
information on the UT fit is used (All) or all the information except
$\vert V_{ub}\vert$ measurements, neither inclusive nor exclusive
(All[no semilep]) is taken. In Tab.~\ref{tab:hadronic} we give the
results for $\hat B_K$, $f_{B_{s}} \, \hat B^{1/2}_{B_{s}}$ and $\xi$
for these two cases.  We also give the values of $f_B$ obtained from
this fit, using in addition the lattice value of $\hat B_{B_d}$.  In
the last column of the table we give the lattice values for an easier
comparison with those extracted from the UT fit.
\begin{table*}[h]
\begin{center}
\begin{tabular}{@{}ccccc}
\hline\hline  
    Parameter     &     All         &  All[no semilep] &  Lattice  \\   \hline
$\hat B_K$        &  $0.94\pm0.17$  &  $0.88\pm 0.13$  & $0.79\pm 0.04\pm 0.08$\\ \hline
$f_{B_{s}} \, \hat B^{1/2}_{B_{s}}$ (MeV)  
                  &  $257\pm6$      & $259\pm 6$       & $262\pm 35$\\ \hline
$\xi$             &  $1.06\pm0.09$  & $1.13 \pm 0.08$  & $1.23 \pm 0.06$ \\ \hline
$f_{B} $ (MeV)    &  $217\pm19$     & $202\pm 16$      & $189\pm 27 $\\ \hline
$f_{B_s} $ (MeV)  &  $227\pm 9$     & $229 \pm 9$    & $230\pm 30$\\ \hline
\hline
\end{tabular} 
\end{center}
\caption {\it Comparison of determinations of the hadronic  parameters
  from the constraints on the angles $\alpha$, $\beta$, and $\gamma$ 
  and  $\vert V_{ub}\vert$ from semileptonic decays (All) or using only the \utangles\ but not 
  the semileptonic decays (All[no semilep]). 
\label{tab:hadronic}}
\end{table*}
\par
\begin{figure}[htb!]
\begin{center}
\includegraphics*[width=0.45\textwidth]{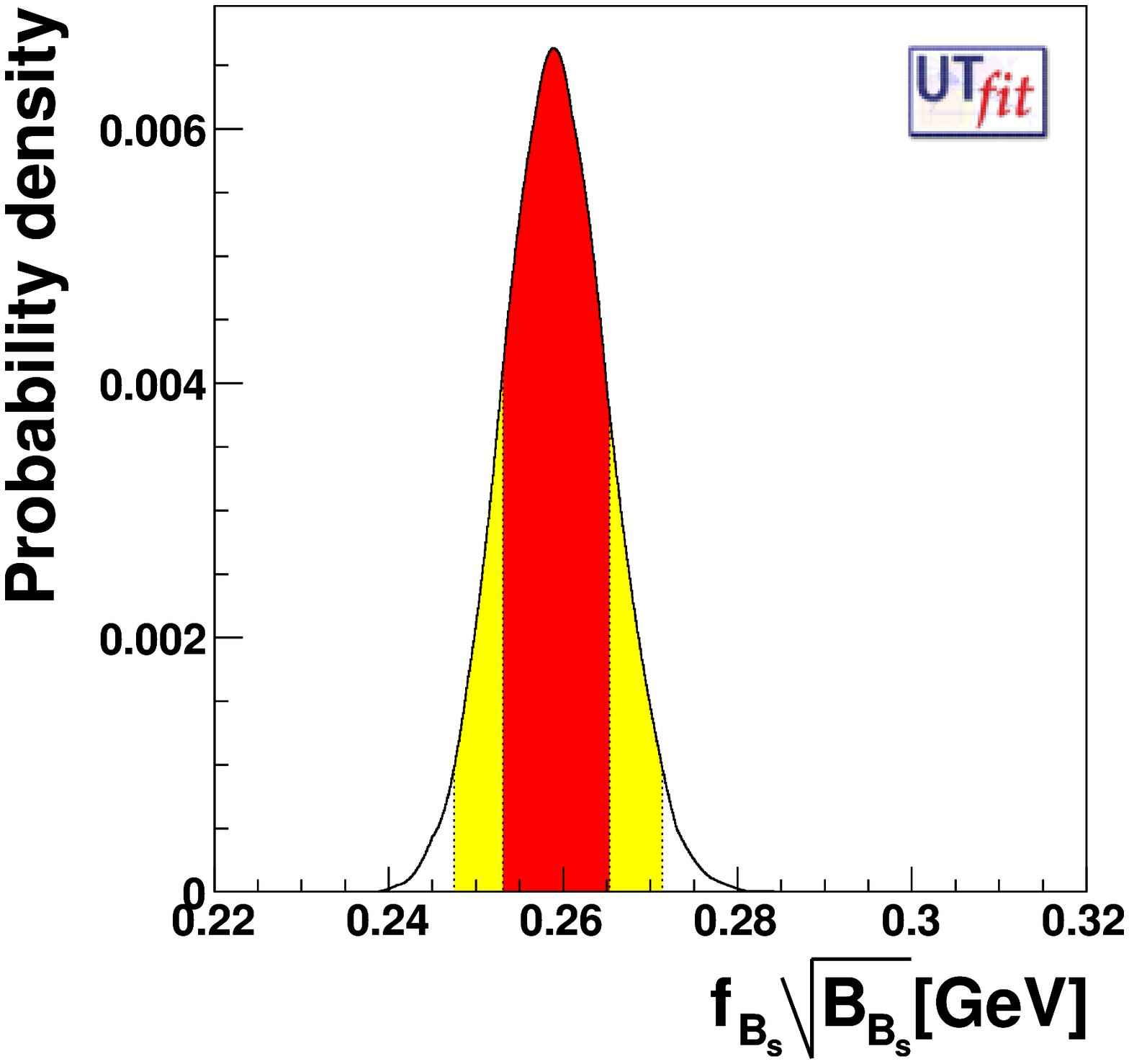}
\includegraphics*[width=0.45\textwidth]{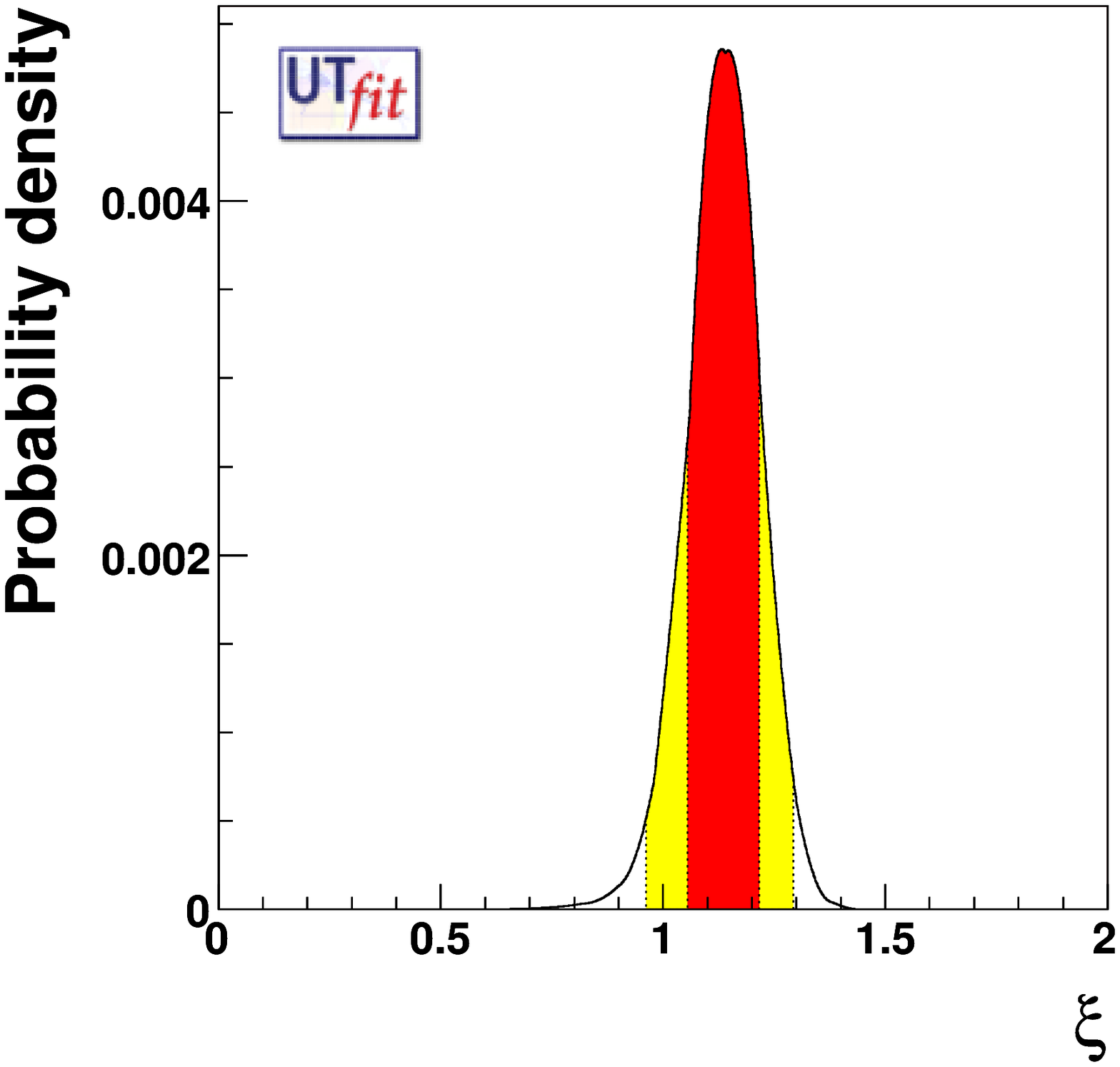} \\
\includegraphics*[width=0.45\textwidth]{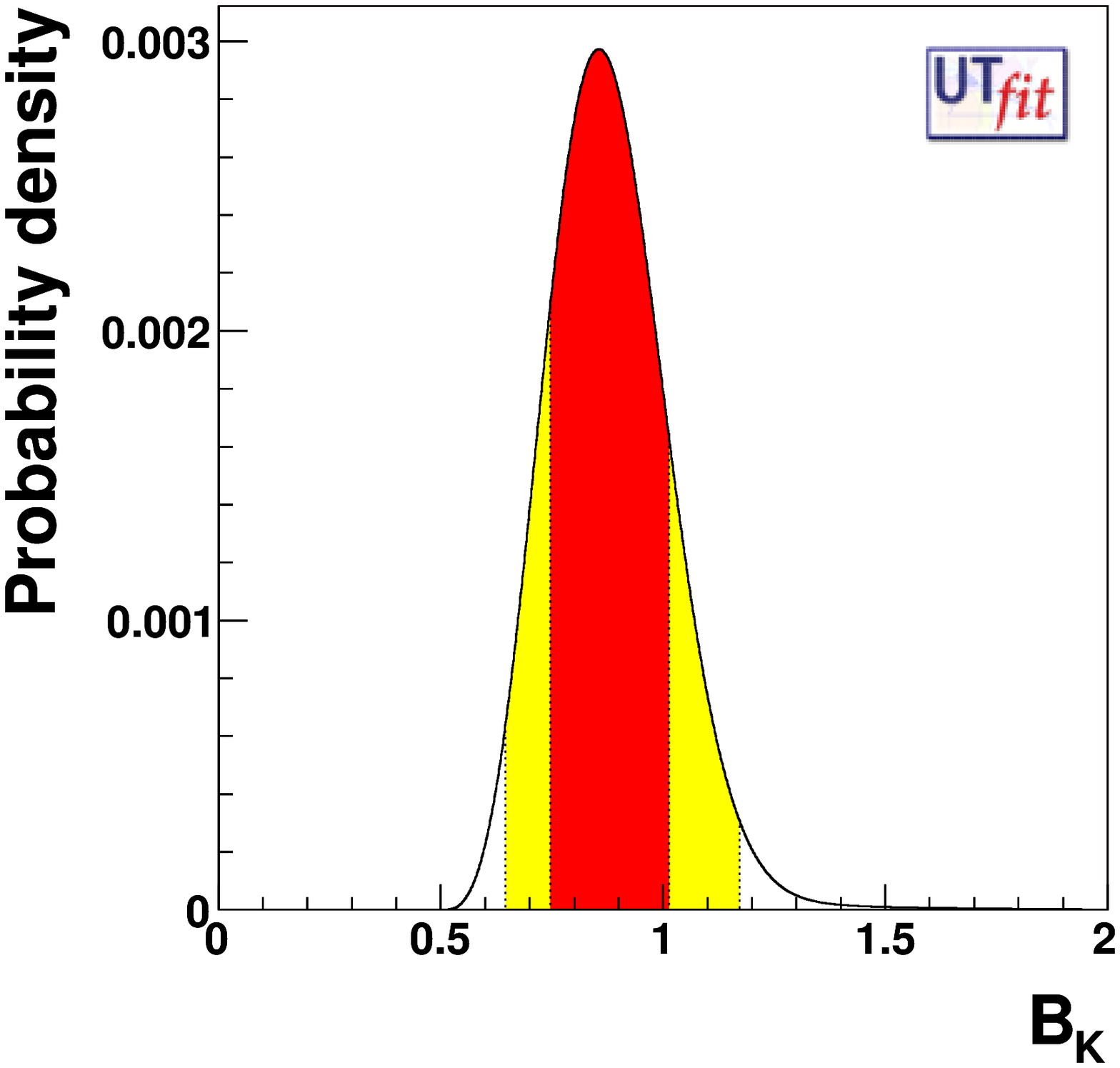}
\includegraphics*[width=0.45\textwidth]{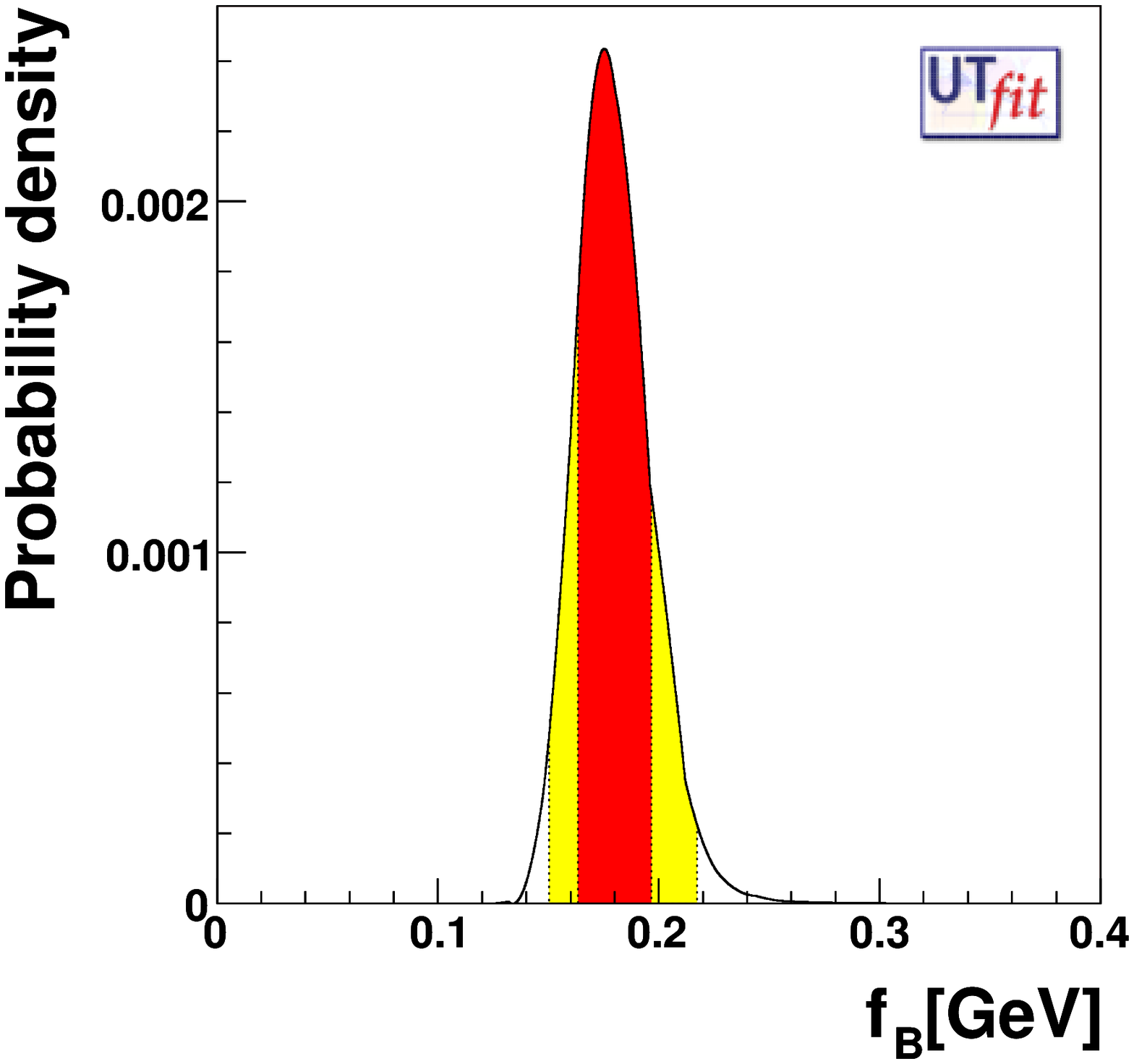}
\caption{%
  \textit{Determination of $f_{B_s}\sqrt{\hat B_s}$ (top-left), $\xi$ (top-right), $\hat B_K$ (bottom-left)
          and $f_{B}$ (bottom-right) 
          obtained from the other UT constraints, using the angles information without using the  semileptonic decays.}}
\label{fig:latticeparam_noVub}
\end{center}
\end{figure}

 We observe a better agreement with lattice calculations when
$\vert V_{ub}\vert$ measurements are not included. Since the
constraint provided by $\vert V_{ub}\vert$ is mainly determined by its
inclusive value, in Figs.~\ref{fig:latticeparam_noVub} we prefer to
give the probability distributions for all the hadronic quantities
considered in this paper ($\hat B_K$, $f_{B_{s}} \, \hat
B^{1/2}_{B_{s}}$, $\xi$, and $f_B$) obtained without using the
semileptonic decays, cfr. the case All[no semilep] in
Tab.~\ref{tab:hadronic}.

\par The value of $f_{B_{s}} \, \hat B^{1/2}_{B_{s}}$ from the UTfit
is essentially independent of $\vert V_{ub}\vert$ and in good
agreement with the lattice prediction (which has, at present, a large
uncertainty).  It is also interesting to extract the value of
$f_{B_{s}}$ using the lattice value of $\hat B_{B_{s}}$, which we take
equal to $\hat B_{B_{d}}$. Using all the constraints we obtain
$f_{B_{s}}=227 \pm 9$~MeV.  The central value is sensibly smaller than
the result predicted by the HPQCD collaboration~\cite{Gray:2005ad},
$f_{B_{s}}=259 \pm 32$~MeV, although compatible within the
uncertainties, and closer to other quenched or partially quenched
results~\cite{hashimoto}.  We believe that other unquenched
calculations of the $f_{B_{s}}$, with different lattice formulations,
are necessary to pin down the lattice uncertainties and make a
meaningful comparison with the ``experimental'' number.  The same holds
true for $f_B$, for which ref.~\cite{Gray:2005ad} quotes a value
larger than many other lattice determinations.\footnote{It is also
  interesting to compare the result of the fit with QCD sum rules
  calculations of the decay constants. For example, ref.~\cite{jamin}
  quotes $f_B=210 \pm 19$ MeV and $f_{B_s}=244 \pm 21$ MeV.}

\begin{figure}[htb!]
\begin{center}
\includegraphics*[width=0.45\textwidth]{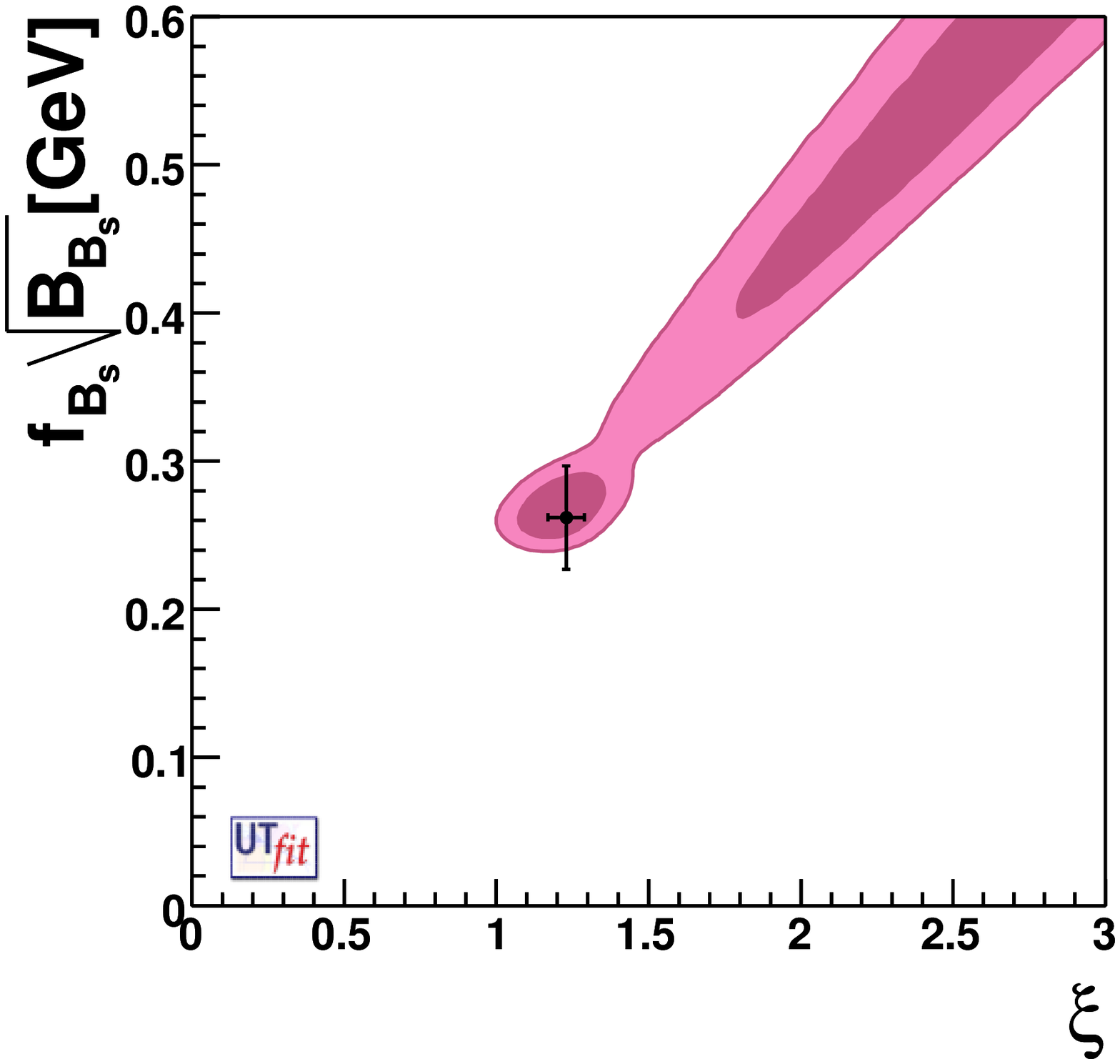}
\includegraphics*[width=0.45\textwidth]{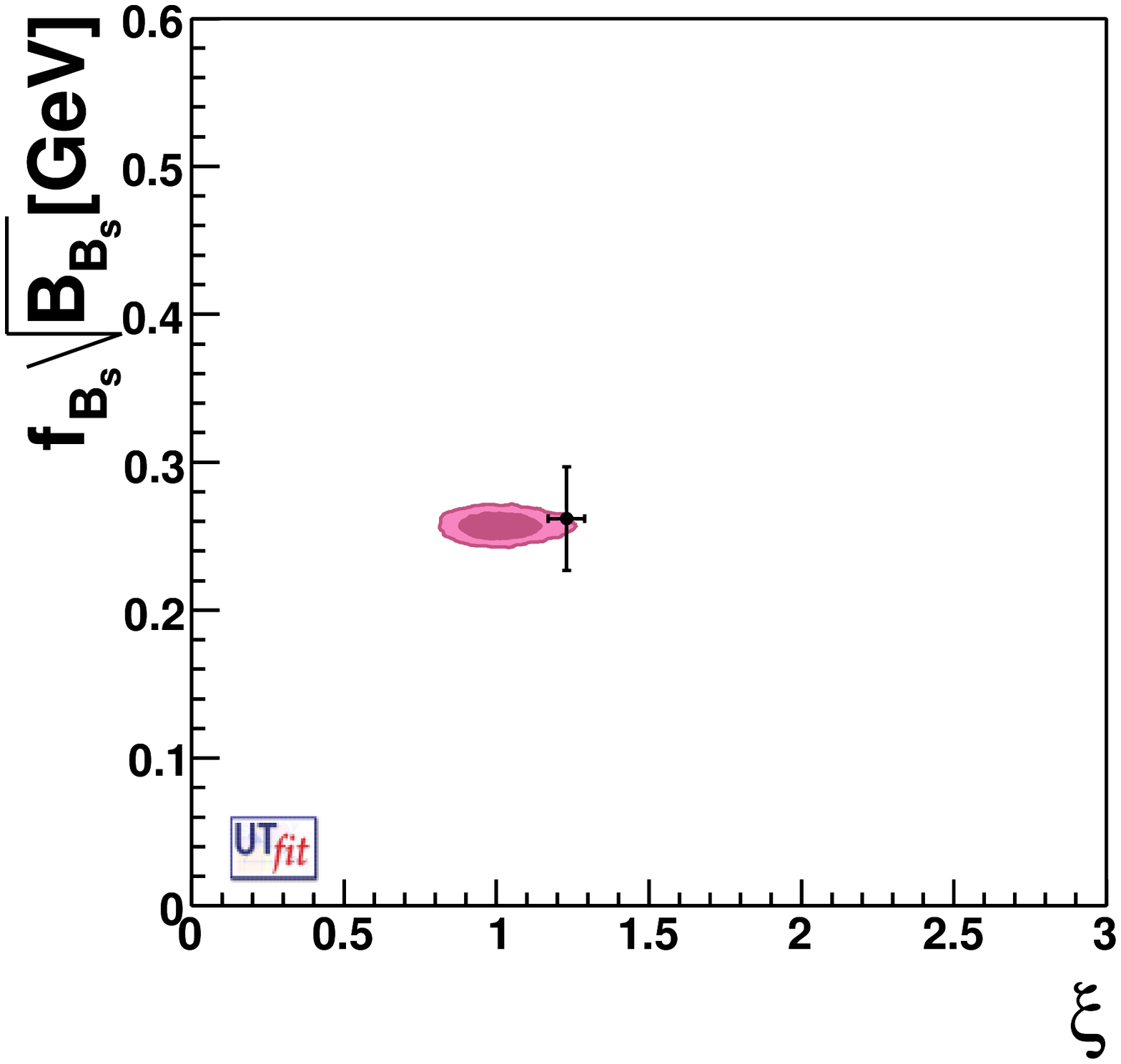}
\caption{%
  \textit{Constraint in the $f_{B_s}\sqrt{\hat B_s}$ vs. $\xi$ plane, using the \utangles\ 
          result for the CKM matrix and the experimental information on $\Delta m_d$
          and $\Delta m_s$. The plot on the right (left) gives the available 
          constraint using  the CDF measurement of  $\Delta m_s$ (the upper bound
          before the CDF measurement). The error bars show  the results from lattice QCD calculations.
\label{fig:2dlattice}}}
\end{center}
\end{figure}

In Figs.~\ref{fig:2dlattice} we show the allowed probability regions
in the $f_{B_{s}} \, \hat B^{1/2}_{B_{s}}$ vs. $\xi$ plane, before and
after the new measurement of $\Delta m_s$.  Before having such input,
we could not put an upper bound on $\xi$ since only the lower limit on
$\Delta m_s$ was available. Now, thanks to the precision of the CDF
determination, the value of $\xi$ is strongly constrained.  This
proves that the CDF measurement of $\Delta m_s$ represents a
substantial progress, not only for the UT analysis, but also for our
knowledge of the hadronic parameters.

The phenomenological extraction of the hadronic parameters and the
comparison with lattice results assumes the validity of the SM and it
is meaningful in this framework only. A similar strategy could be
followed in any given extension of the SM when enough experimental
information is available.  In general, however, a model-independent UT
analysis beyond the SM cannot be carried out without some ``a priori''
theoretical knowledge of the relevant hadronic parameters.  For this
reason the error in the calculation of the hadronic matrix elements
affects the uncertainties in the determination of the NP
parameters~\cite{UTNP,citarenoinpreparation}.

\section{Conclusions}  
The recent precise determination of $\Delta m_s$ by the CDF
Collaboration allows a substantial improvement of the accuracy of the
UT fit. Thanks to this new measurement, and to the determination of
the leptonic branching fraction $BR(B \to \tau \nu_\tau)$ by Belle, we
have shown that it is possible to extract from experiments the value
of the relevant hadronic parameters, within the Standard Model.  It is
remarkable that the measurement of $\Delta m_s$, combined with all the
information coming from the UT fit, allows the determination of
$f_{B_{s}} \, \hat B^{1/2}_{B_{s}}$ with an error of 6~MeV ($f_{B_{s}}
\, \hat B^{1/2}_{B_{s}}= 257\pm6$~MeV) and of $f_{B_{s}}$ with an
error of 9~MeV ($f_{B_{s}}=227\pm 9$~MeV).  The accuracy in the
determination of $\xi$ suffers instead from the strong correlation
that it has with the value and uncertainty on $\vert V_{ub}\vert$.

The only exception to the general consistency of the fit is given by
the inclusive semileptonic $b \to u$ decays the analysis of which
relies on the parameters of the shape function.  We observed that the
present determination of $\vert V_{ub}\vert$, using inclusive methods,
is disfavoured by all other constraints at the 2.5$\sigma$ level.
This can come either from the fact that the central value of $\vert
V_{ub}\vert$ from inclusive decays is too large, or from the smallness
of the estimated error, or both.  Moreover the problem has been
recently worsened by the decrease of the value of $\sin(2 \beta)$
determined by the direct measurements. We think that it is worth
investigating whether the theoretical uncertainty of the inclusive
analysis has been realistically estimated.

$\vert V_{ub}\vert$ from exclusive decays has still large
uncertainties and the only conclusion that we may draw is that an
effort must be done for a substantial improvement of the theoretical
and experimental accuracy for this quantity.

In the future, a confirmation of the results presented in this paper
with smaller errors might reveal the presence of NP in the generalized
UT analysis~\cite{citarenoinpreparation,gino}. Before claiming such
results, a better accuracy on the determination of $\vert V_{ub}\vert
$ is however needed.

\section*{Acknowledgements}
We thank G. D'Agostini and R.~Faccini for informative discussions.
This work has been supported in part by the EU network ``The quest for
unification'' under the contract MRTN-CT-2004-503369.

\end{document}